\documentclass{article}
\usepackage{amsmath,amssymb,epsfig}

\newcommand{\as}{\alpha_{\mathrm{s}}}

\newcommand{\dif}{\mathrm{d}}
\newcommand{\GeV}{\;\mathrm{GeV}}
\newcommand{\kt}{{\boldsymbol k}}
\newcommand{\ord}[1]{\mathcal{O}\left(#1\right)}
\newcommand{\pt}{{\boldsymbol p}}
\newcommand{\vm}[1]{\langle #1 \rangle}

\begin{document}
\title{Improved theoretical description of Mueller-Navelet jets at LHC
\thanks{Presented at  the Low x workshop, June 13-18 2017, Bari, Italy}%
}
\author{
  D.\ Colferai\footnote{colferai@fi.infn.it} \footnote{Speaker.}\\
  {\small Dipartimento di Fisica ed Astronomia, Universit\`a di Firenze and INFN
    Firenze}
  \smallskip\\
  F.\ Deganutti\footnote{fedeganutti@ku.edu}\\
  {\small Department of Physics and Astronomy, Kansas University}
  \smallskip\\
  A.\ Niccoli\footnote{alessandro.niccoli@unifi.it}\\
  {\small Dipartimento di Fisica ed Astronomia, Universit\`a di Firenze}
  \smallskip\\
}
\date{}
\maketitle
\begin{abstract}
  We present a method for improving the phenomenological description of
  Mueller-Navelet jets at LHC, which is based on matching the BFKL resummation
  with fixed order calculations.  We point out the need of a consistent
  identification of jets between experimental measurements and theoretical
  descriptions.  We hope as well to motivate an extensive analysis of MN jets at
  LHC in run 2.\\
  ~
  \\
  PACS number(s): 12.38.Cy, 13.87.-a
\end{abstract}

\section{Outline}

Mueller-Navelet (MN) jets~\cite{MNJ} are one of the most suitable observables
for investigating QCD in the high energy limit, where QCD has a new kind of
dynamics, showing power-like behaviour of the amplitudes and cross-sections.  In
the first part of the talk we shall review the theoretical description of MN
jets within the BFKL framework~\cite{BFKL}.  Then we shall show a part of the
experimental analysis by the CMS collaboration and its comparison with BFKL and
also with MonteCarlo (MC) predictions.

Since such comparisons are not satisfactory, some improvements have been
proposed by some groups. We shall present our proposals of improvement:
\begin{itemize}
\item firstly, we shall point out that so far theoretical descriptions are not
  consistent with experimental analysis, regarding the jet selection; we shall
  show how to modify the theoretical description in a way consistent with
  experimental analysis;
\item next we shall present our method based on matching the resummed BFKL
  description with the fixed next-to-leading order (NLO) calculations, and we
  shall show some preliminary results.
\end{itemize}

\section{Mueller-Navelet jets: theoretical description}

MN jets are inclusive events with a pair of jets having large rapidity
separation and comparable transverse momenta~\cite{MNJ}.  The theoretical
description of MN jets is based on a double factorization formula
\begin{align}\label{fact}
  \frac{\dif\sigma_{AB}(s)}{\dif J_1 \dif J_2}
  &= \sum_{a,b}\int_0^1 \dif x_1 \dif x_2 \int\dif^2\kt_1\,\dif^2\kt_2 \;
  f_{a/A}(x_1) V_a(x_1,\kt_1;J_1) \times \nonumber \\
  &\quad G(x_1 x_2 s, \kt_1, \kt_2) V_b(x_2,\kt_2;J_2) f_{b/B}(x_2) \;,
 \quad \dif J_i \equiv \dif y_i \,\dif E_i \,\dif\phi_i \;,
\end{align}
where $J$ collects all jet variables: rapidity $y$, transverse energy $E=|\pt|$
and azimuth $\phi$.

On the external side, we have the usual collinear factorization which expresses
the hadronic differential cross-section as a convolution of partonic
distribution functions $f$ and a hard-scattering partonic cross-section. In
turn, the partonic cross-section at high energy can be expressed by means of a
$\kt$-dependent factorization formula.  In the middle there is the so-called
gluon Green function (GGF) $G$, which represents the sum of all ladder diagrams
with reggeized gluon exchanges, and obeys the BFKL equation~\cite{BFKL}, whose
integral kernel represents one rung of the ladder and is computable in
perturbation theory.  The impact factors $V$ describe the coupling of the
Reggeized gluon with the external particles. In this case of incoming parton and
outgoing jet, the impact factors are called jet vertices.

At leading-logarithmic (LL) level, the jet is trivial, being simply identified
with the scattered parton, since the LL kinematics is characterized by large
rapidity gaps among emitted particles.  At next-to-leading logarithmic (NLL)
level, Bartels, Vacca and myself proved a factorization formula with the same
structure~\cite{BCV02}. Both the GGF and jet vertices receive NLL correction,
and in this case the jets can have a non-trivial structure. In particular they
depend on the jet resolution $R$ and on the jet algorithm.

Now, with LHC, we can test these ideas.  In fact, few years ago Schwennsen,
Szymanowski, Wallon and myself made a careful study of MN jets for the design
energy at 14 TeV~\cite{CSSW10}.  We noticed that the NLL corrections to the jet
vertices are sizeable and as important as those to the GGF; furthermore the
ensuing predictions were definitely different from those based on MonteCarlos
with fixed-order matrix elements.  Therefore, we have an handle for finding
signals of the high-energy dynamics we are looking for.

\section{Experimental analysis by CMS}

From the experimental side, in 2012 CMS published an analysis of MN jets from
data collected during the 7 TeV run~\cite{CMSmnj}.  They have analysed the
distribution of the azimuthal angle $\Delta\phi\equiv\phi_1-\phi_2-\pi$ between
the two jets. It is zero when they are emitted back-to-back, as in lowest-order
perturbation theory, while it can be different from zero when additional
radiation is present.

Since BFKL predicts an amount of radiation that is different (typically larger)
than fixed order calculations, one could expect to reveal the signal of BFKL
dynamics from decorrelation measurements.  CMS measured the
average value of $\cos(\Delta\phi)$ and higher angular moments
$C_m\equiv\int_0^{2\pi}\dif\Delta\phi\;\frac{\dif^2\sigma}{\dif\Delta\phi\,\dif
  Y}\cos(m\Delta\phi)$, as well as their ratios $C_m/C_n$. The advantage of
these observables relies in the partial cancellation of various sistematic
errors like PDF and scale uncertainties.

\begin{figure}[ht]
  \centering
  \includegraphics[width=0.85\textwidth]{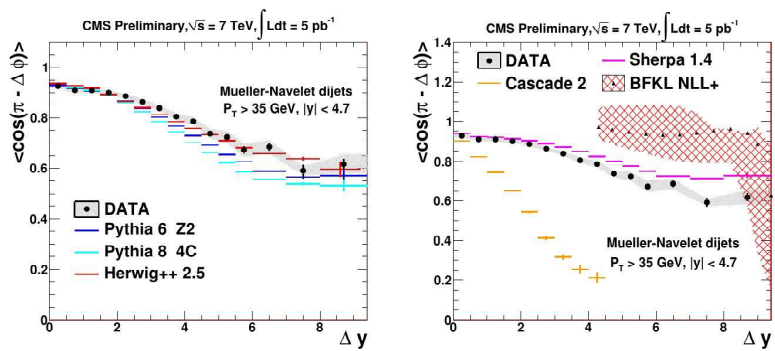}
  \caption{The MN jet angular coefficient $\vm{\cos(\Delta\phi)}$ measured by CMS.}
  \label{f:cms1}
\end{figure}

Fig.~\ref{f:cms1} shows the CMS measurements of the average of
$\cos(\Delta\phi)$, i.e. $C_1/C_0$, at various rapidity differencies
$Y\equiv y_1-y_2$.  Data are compared with both MC and BFKL predictions.  As
expected, with increasing $Y$ there is more radiation and thus more
decorrelation. However BFKL, represented by the red hatched band, shows less
decorrelation, contrary to expectations. Some MC are close to data, but others
are not and NLL BFKL is definitely off, despite the rather large uncertainty
band due essentially to scale variations.

The situation is similar for higher angular moments like $C_2$.  Surprisingly,
the ratio $C_2/C_1$ is in perfect agreement with BFKL, while MCs don't agree
with data.  In practice, Neither BFKL NLL nor fixed order MC give a satisfactory
description of data yet; in addition BFKL at NLL still suffers from large scale
uncertainties of the order of 10$\div$15\%.

In order to improve the BFKL description, various optimization procedure were
used. It was proposed in~\cite{DSW13} to tame the large scale dependence of BFKL
by fixing the renormalization scale according to the BLM procedure~\cite{BLM}.
The underlying idea is to effectively include higher order corrections which are
missing in NLL BFKL.  However, as shown in~\cite{CIMP14}, the obtained results
are quite different for the various prescriptions, and eventually only BLM does
a fairly good job.

To include higher order corrections is probably needed, but to some extent it is
an arbitrary procedure. On the other hand, there is a rigorous improvement that
can be done with the actual knowledge of QCD, that is matching NLL BFKL with
perturbative next-to-leading order (NLO) (and possibly NNLO) calculations, and
this is the path that we shall follow.

\section{Correction of the jet selection algorithm}

Actually, during our recent studies, we realized that the definition of jet
vertices proposed in~\cite{BCV02}, checked in 2011 by Caporale, Ivanov, Murdaca,
Papa and Perri~\cite{CIMPP11}, and used until now in the description of MN jets,
is not equivalent to the event selection adopted in the measurement by CMS. Let
me first discuss this issue of event selection, and afterwards return to the
proposal of improvement through the matching procedure.

\begin{figure}[ht]
  \centering
  \includegraphics[width=0.5\textwidth]{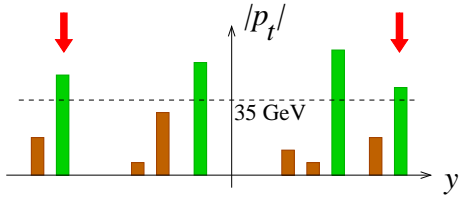}
  \caption{Schematics of jet selection by CMS.}
  \label{f:CMSsel}
\end{figure}

This mismatch can be easily understood with an example:
\begin{itemize}
\item in the experimental analysis, particles --- represented in
  fig.~\ref{f:CMSsel} by bars whose position indicates rapidity and whose height
  indicates transverse energy $|\pt|$ --- are first clustered into jets;
\item then one considers only jets with $|\pt|$ above some threshold $E_0$, in
  that case 35~GeV, highlighted in green in fig.~\ref{f:CMSsel};
\item finally, the tagged jets (i.e., the MN jets) are the farthest in rapidity,
  as indicated by the red arrows.
\end{itemize}

This is not exactly what was prescribed in the original definition of jet
vertices in NLL approximation. Let me sketch the procedure for determining the
vertex in the forward region, namely with positive rapidity:
\begin{itemize}
\item the first step is to choose a set of numbers for the jet variables
  ($y,|\pt|,\phi,R)$;
\item next, for each event, the jet clustering algorithm is applied to the two
  partons with largest rapidities; the cross-section receives contribution only if
  there is a jet with the specified variables;
\item if such jet is formed by both partons, the probability of this event
  contributes to the NL jet vertex;
\item instead, if the jet contains only one parton, we look at the other
  parton:
  \begin{itemize}
  \item if it is found in the fragmentation region of incoming hadron, this
    contribution is collinear singular, hence a PDF correction;
  \item if it is a gluon in central region, this event contributes to the GGF;
  \item in all other cases, the events contribute to the jet vertex correction.
  \end{itemize}
\end{itemize}

The important point to notice is that, in this last case, the parton outside the
tagged jet can be hard ($\pt > 35\GeV$) and can also be emitted at rapidity
$y>y_J$.  At the hadronic level, this parton gives rise to a jet. This means
that we are allowing jets with rapidity greater than that of the MN jet, in
contrast to the experimental selection of events previously described.

In other words, if the two partons 1,2 with largest rapidity are not clustered
into a single jet, and both of them have $\pt > 35 \GeV$, this event contributes
twice to the MN jet cross-section: one time with $J=p_1$, the other time with
$J=p_2$.  On the contrary, in the CMS experimental analysis, this event
contributes only once, the jet $J$ being identified always with the parton
having largest rapidity.

\begin{figure}[ht]
  \centering
  \includegraphics[width=0.4\linewidth]{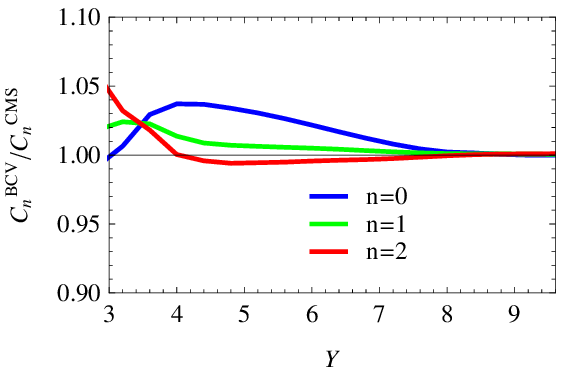}
  \hspace{0.07\linewidth}
  \includegraphics[width=0.4\linewidth]{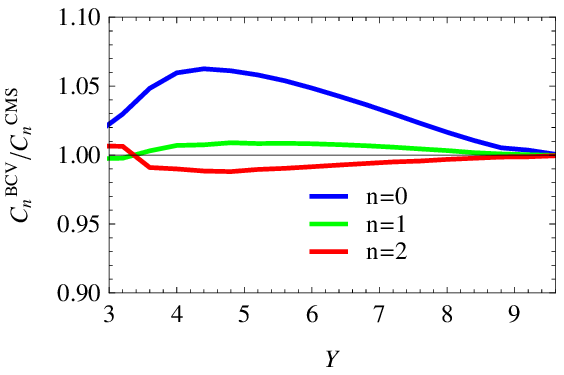}
  \caption{Relative difference of the MN jet cross-sections between the two
    different choices of event selections described in the text, for 3 values of
    the angular index $n$. Left: $\sqrt{s}=7$ TeV; right: $\sqrt{s}=13$ TeV.}
  \label{f:deltaSigma}
\end{figure}

Conceptually, this is a big difference.  However, we have to remember that,
since the MN jets have typically a large rapidity distance, it is rather
unlikely to have the emission of another parton at even larger absolute
rapidities. For this reason, we don't expect big differences between the two
procedures. In practice we find differences of the order of 4\% at intermediate
rapidities ($Y\simeq 3\div 5$) and much smaller at larger rapidities.
Fig.~\ref{f:deltaSigma} shows the relative difference between the NLL BFKL
predictions with the original NLL jet vertices (BCV) and the one with modified
jet selection that prevents the emission of a parton with $\pt > 35\GeV$ and
$y>y_J$ (CMS).  We note that the relative difference almost doubles at 13 TeV,
and cannot be neglected.
Let me just mention that it could be possible to modify the experimental
analysis of MN jets in order to comply with the original proposal of NLO jet
vertices, but we think that it is better to modify the theoretical prescription
by requiring the absence of jets with $\pt>35\GeV$ and $y>y_J$.

\section{The matching procedure}

Finally, we want to present the matching procedure.  We believe that exploiting
all the actual knowledge of perturbation theory should produce more reliable
results and improve the description of data, with reduced scale uncertainties.
Our hope is also to improve the estimate of cross-section, and not just the
azimuthal decorrelation. In this case, we think that fixed NLO and probably NNLO
corrections must be fully taken into account.

The matching procedure is rather standard, though tricky at some point, as you
will see.  In practice, we add to BFKL the full perturbative NLO result, and
then subtract the $\ord{\as^3}$ part already included in BFKL, in order to avoid
double counting.  The implementation is still work in progess, therefore we can
show only preliminary results of central values, without error estimate yet.
However, we can learn an important lesson for future phenomenological analysis.

\begin{figure}[ht]
  \centering
  \includegraphics[width=0.9\textwidth]{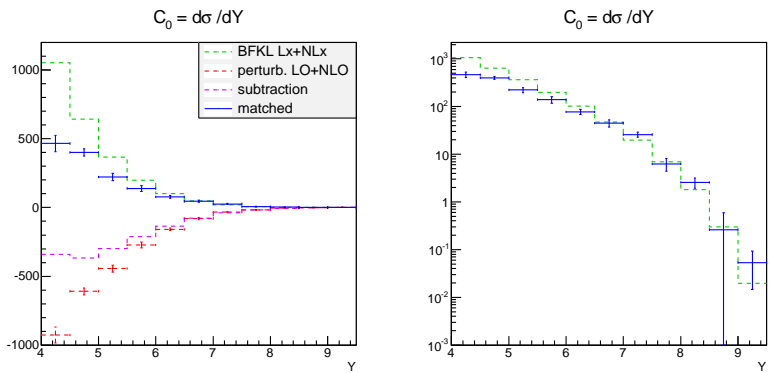}
  \caption{MN jet differential cross-section: NLL BFKL (green), fixed NLO (red),
    common part (purple), matched cross-section (blue). On the left we use a
    linear scale for the cross-section; on the right a logarithmic scale.}
  \label{f:matching}
\end{figure}

Let me start with the computation of the differential cross-section with respect
to the rapidity difference of the two jets. Here the center-of-mass energy is 7
TeV and we impose symmetric cuts on the two jet transverse momenta, as in the
CMS analysis.  In fig.~\ref{f:matching} we show the NLL BFKL prediction in
green, and compare it with the fixed NLO result in red. The latter turns out to
be negative! Furthermore, after several days of computing time, MC-integration
errors are still large, due to the slow convergence of the integrand.  However,
also the double-counting subtraction (purple) is negative. Since the difference
between red and purple is moderate, the matched cross-section (blue) receives a
moderate correction and remains positive.
An analogue behaviour is shared by the first azimuthal coefficient $C_1$. The
matching procedure provides moderate but definitely sizeable corrections, in
particular at intermediate values of $Y$.

Let me briefly discuss the issue of negative cross-section and instability of
jet cross-section with symmetric cuts. It is well known that jet cross-sections
at NLO are very sensitive to the asymmetry parameter $\Delta$ which measures the
imbalance between the $\pt$-cuts on the two jets. Despite the fact that NLO
cross-sections are finite for all values of $\Delta$, they are affected by a
$\Delta\log(\Delta)$ singularity of collinear origin, which provides an infinite
derivative at $\Delta=0$, that is with symmetric jet cuts, and even negative
cross-sections for small $\Delta$-values, as we have seen.

However, and analogous singularity occurs in the subtraction term, as can be
seen in the expansion of the BFKL cross-section.  It turns out that in the
matching procedure such singular terms cancel out to a large extent, so that the
matching is quite safe.

\begin{figure}[ht]
  \centering
  \includegraphics[width=0.9\textwidth]{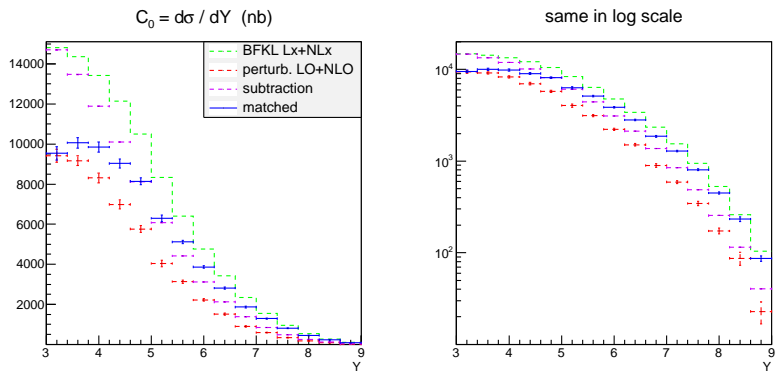}
  \caption{MN jet differential cross-section with cut on the average jet
    transverse energy $\frac12(|\pt_1|+|\pt_2|)>50\GeV$.}
  \label{f:ptsum}
\end{figure}

Actually, this problem could be avoided by using asymmetric cuts on the jet
transverse energies, but this introduces an unpleasant asymmetry.  Fortunately,
there is another way to avoid (or at least to reduce) this problem while keeping
symmetry between the two jets, that is to impose a cut on the average jet
transverse energy: $\frac12(|\pt_1|+|\pt_2|)>E_{\mathrm{cut}}$.  In this case,
also the fixed-order cross-section is well behaved, and the whole procedure is
more stable than the previous one, as shown in fig.~\ref{f:ptsum}.  The same is
true for the $C_1$ angular coefficient.

For this reason, we strongly suggests experimentalists to perform MN jets
analysis with a cut on the average of the jet transverse energies, as already
done in other jet analysis. This allows smaller theoretical uncertainties so
that MN jets become a better tool for finding evidence of BFKL dynamics, which
at present energies is still competing with fixed order contributions, even at
LHC.

\section{Conclusions}

To conclude, MN jets are a good observable for demonstrating the presence of
BFKL dynamics at high energy, with yet open questions to be answered.
Fixed order MC and NLL BFKL descriptions are quite different, in some cases
close to data, but the overall agreement is not good.

We observed that the theoretical formulation of jet vertices has to be modified
in order to comply with the experimental analysis that identifies as MN jets
those with the largest rapidity separation.  We propose an improved theoretical
description of MN jets based on matching BFKL resummed calculations with
fixed-order ones. We have computed various observables and we will provide a
full analysis with estimates of theoretical uncertainties.

We think that a precise experimental analysis from the 13 TeV run would be very
valuable for the study of QCD at high energies. In this respect, we strongly
suggest experimentalists of LHC to carry out such analysis by imposing
different cuts on the jet transverse momenta, for instance a cut on the average
transverse energy of jets.


\bibliographystyle{apsrev4-1}

\begin{thebibliography}{99}

\bibitem{MNJ}
  A.~H.~Mueller and H.~Navelet,
  Nucl.\ Phys.\ B {\bf 282} (1987) 727.
  doi:10.1016/0550-3213(87)90705-X

\bibitem{BFKL}
  V.~S.~Fadin, E.~A.~Kuraev and L.~N.~Lipatov,
  Phys.\ Lett.\ B {\bf 60} (1975) 50;
  E.~A.~Kuraev, L.~N.~Lipatov and V.~S.~Fadin,
  Sov.\ Phys.\ JETP {\bf 44} (1976) 443
   [Zh.\ Eksp.\ Teor.\ Fiz.\  {\bf 71} (1976) 840]
   [Erratum-ibid.\  {\bf 45} (1977) 199];
  E.~A.~Kuraev, L.~N.~Lipatov and V.~S.~Fadin,
  Sov.\ Phys.\ JETP {\bf 45} (1977) 199
   [Zh.\ Eksp.\ Teor.\ Fiz.\  {\bf 72} (1977) 377];
  I.~I.~Balitsky and L.~N.~Lipatov,
  Sov.\ J.\ Nucl.\ Phys.\  {\bf 28} (1978) 822
   [Yad.\ Fiz.\  {\bf 28} (1978) 1597].

\bibitem{BCV02}
  J.~Bartels, D.~Colferai and G.~P.~Vacca,
  Eur.\ Phys.\ J.\ C {\bf 24} (2002) 83
  [hep-ph/0112283];
  J.~Bartels, D.~Colferai and G.~P.~Vacca,
  Eur.\ Phys.\ J.\ C {\bf 29} (2003) 235
  [hep-ph/0206290].

\bibitem{CSSW10}
  D.~Colferai, F.~Schwennsen, L.~Szymanowski and S.~Wallon,
  JHEP {\bf 1012} (2010) 026
  [arXiv:1002.1365 [hep-ph]].

\bibitem{CMSmnj}
  CMS Collaboration [CMS Collaboration],
  CMS-PAS-FSQ-12-002.

\bibitem{DSW13}
  B.~Duclou\'e, L.~Szymanowski and S.~Wallon,
  JHEP {\bf 1305} (2013) 096
  [arXiv:1302.7012 [hep-ph]].

\bibitem{BLM}
  S.~J.~Brodsky, G.~P.~Lepage and P.~B.~Mackenzie,
  Phys.\ Rev.\ D {\bf 28} (1983) 228.
  doi:10.1103/PhysRevD.28.228

\bibitem{CIMP14}
  F.~Caporale, D.~Y.~Ivanov, B.~Murdaca and A.~Papa,
  Eur.\ Phys.\ J.\ C {\bf 74} (2014) no.10,  3084
   Erratum: [Eur.\ Phys.\ J.\ C {\bf 75} (2015) no.11,  535]
  doi:10.1140/epjc/s10052-014-3084-z, 10.1140/epjc/s10052-015-3754-5
  [arXiv:1407.8431 [hep-ph]].

\bibitem{CIMPP11}
  F.~Caporale, D.~Y.~Ivanov, B.~Murdaca, A.~Papa and A.~Perri,
  JHEP {\bf 1202} (2012) 101
  doi:10.1007/JHEP02(2012)101
  [arXiv:1112.3752 [hep-ph]].

\end{thebibliography}

\end{document}